# High-performance K-means Implementation based on a Simplified Map-Reduce Architecture

Zhehao Li, Jifang Jin, Lingli Wang
State Key Laboratory of ASIC & System, Fudan University
Shanghai, China
{lizh13, jfjin14, llwang}@fudan.edu.cn

## ABSTRACT
**The *k*-means algorithm is one of the most common clustering algorithms and widely used in data mining and pattern recognition. The increasing computational requirement of big data applications makes hardware acceleration for the *k*-means algorithm necessary. In this paper, a simplified Map-Reduce architecture is proposed to implement the *k*-means algorithm on an FPGA. Algorithmic segmentation, data path elaboration and automatic control are applied to optimize the architecture for high performance. In addition, high level synthesis technique is utilized to reduce development cycles and complexity. For a single iteration in the *k*-means algorithm, a throughput of 28.74 Gbps is achieved. The performance shows at least 3.93x speedup compared with four representative existing FPGA-based implementations and can satisfy the demand of big data applications.**

## CCS Concepts
• **Hardware➜Integrated circuits➜Reconfigurable logic and FPGAs➜Hardware accelerators.**

## 1. INTRODUCTION
The *k*-means algorithm is an unsupervised clustering algorithm to partition the input samples into *k* clusters, so that samples within a cluster share similar attributes, while dissimilar samples are grouped into different clusters [1]. The algorithm is widely applied to applications ranging from data mining, pattern recognition to bioinformatics.

Albeit powerful, the *k*-means algorithm becomes time-consuming as the input sample set grows large, rendering hardware acceleration necessary, especially for big data applications.

Field Programmable Gate Array (FPGA) is characterized by its natural feature of parallelism, which makes it an applicable platform for exploiting the algorithmic parallelism and accelerating the *k*-means algorithm. Over the last decade, several hardware designs for the *k*-means algorithm were proposed and implemented on the FPGA. Despite displaying respectable performance, most of these implementations can hardly satisfy the demand for computing power and precision required by large-scale clustering tasks nowadays. Among these implementations, some are not optimized for high performance due to their inefficient algorithmic segmentation and data path [1], while others show a lack in precision because of the use of fix-point arithmetic or bitwidth truncation [2][3].

To solve the problems above, a simplified Map-Reduce architecture is proposed to implement the *k*-means algorithm on an FPGA in this paper. The main contributions of this paper are listed below:

- The *k*-means algorithm is adapted for the proposed simplified Map-Reduce architecture by algorithmic segmentation, so that the intrinsic parallelism of the algorithm can be fully exploited.

- The high-performance and high-precision hardware accelerator for the *k*-means algorithm is developed with high level synthesis (HLS) technique. Stream interfaces and the elaborated system architecture are applied to ensure high-speed data transmission between the memory and the accelerator.

- The host programs for task scheduling and data management in traditional Map-Reduce frameworks are implemented in hardware circuits to reduce the communication overhead between the host and the hardware accelerator.

The proposed architecture is implemented on an Xilinx ZC706 Board [4]. Evaluation result shows that a throughput of 28.74 Gigabit per second (Gbps) is achieved for one iteration in the *k*-means algorithm. The performance displays at least 3.93x speedup when compared with four representative existing FPGA-based implementations.

The remainder of this paper is organized as follows: Section 2 gives the background of the *k*-means algorithm and an overview on the related work. Section 3 discusses about design considerations. Section 4 provides details of the implementation. Section 5 shows experiment results. Section 6 concludes the paper.

## 2. BACKGROUND AND RELATED WORK
### 2.1 *K*-means Clustering
The *k*-means algorithm proceeds in iterations until a convergence is reached. At first, cluster centroids are initialized randomly or in heuristic ways. Then during each iteration, the algorithm consists of two steps: sample clustering and cluster centroids updating. In the sample clustering step, distances between each sample and all the cluster centroids are calculated, then each sample is assigned to the nearest cluster and marked with a label that indicates the cluster it belongs to. The distance calculation is based on a distance metrics, such as Euclidean or Manhattan distance. Then in the cluster centroids updating step, the means of samples in each cluster are calculated by accumulation and division to update cluster centroids for use in next iteration. During each iteration, a distortion error is also calculated, which measures the sum of distances between each sample and the cluster centroid to which it has been assigned [5]. The iteration repeats until the change of the distortion error is tiny enough during two consecutive iterations.

## 2.2 Related Work

Early $k$-means implementations on FPGA focus on clustering hyper-spectral and multi-spectral images. Dominique Lavenier [6] designs a systolic array architecture to accelerate the distance calculation in the $k$-means algorithm. The design is evaluated on different FPGA boards for comparison. The work is later improved and implemented on a hybrid processor [7]. A maximum speedup of 11.8x is achieved over a software implementation. Mike Estlick et al. [2] apply algorithmic transformations to map the $k$-means algorithm to the reconfigurable hardware. Selected metrics for distance calculation and bitwidth truncation are used to optimize the performance and reduce the consumption of hardware resource. During evaluation, a speedup of 50x over a software implementation is achieved. All the designs above only implement part of the $k$-means algorithm on hardware, while the rest part is executed on the host.

Venkatesh Bhaskaran [8] is the first to implement a complete $k$-means algorithm on an FPGA. The division is implemented on hardware using dividers from the Xilinx Core Generator. Then Hussain et al. [9] propose a multi-core architecture for the $k$-means algorithm to process microarrays. A 51.7x speedup is achieved over a software implementation when five cores are applied. Besides, several implementations are proposed to improve the performance by algorithmic optimizations. In references [3][10][11], the kd-tree data structure and the triangle inequality are applied to improve the performance of each iteration in the $k$-means algorithm. Despite displaying respectable performance, implementations above can hardly satisfy the demand for high throughput required by large-scale clustering tasks nowadays, due to the insufficient bandwidth of processing elements or the inefficient data path. Moreover, uses of fixed-point arithmetic and bitwidth truncation render low precision of calculation and accuracy of clustering.

To adapt the $k$-means algorithm for large-scale clustering tasks, two widely-used programming models for parallel computing, OpenCL and Map-Reduce, are implemented on reconfigurable hardware. Ramanathan et al. [12] propose an OpenCL-based architecture to accelerate the $k$-means algorithm. The method work-stealing is leveraged for runtime load balancing. Besides, Choi et al. [1] propose a multi-FPGA implementation following the Map-Reduce programming model. The system mainly consists of Mappers and Reducers and each iteration of the $k$-means algorithms is regarded as a Map-Reduce job. The sample clustering is assigned to the Mappers, while the update of cluster centroids is executed in the Reducers. As the numbers of Mappers and Reducers are configurable and samples are stored in the hard-disk rather than the on-chip memory on an FPGA, this implementation provides a practical framework for big data applications. However, both of the implementations above conform too strictly to their corresponding programming models, which are general-purpose and thus contain some unnecessary contents for the $k$-means algorithm. Those unnecessary contents prevent the implementations from achieving satisfying performance. In contrast, a simplified Map-Reduce architecture for the $k$-means algorithm is proposed in this paper. Unnecessary contents such as the shuffle and sort processes and the key/value pair in a traditional Map-Reduce model are removed. The proposed architecture can be optimized to achieve high performance in big data applications.

## 3. DESIGN CONSIDERATIONS

In the $k$-means algorithm, the most computational-intensive part lies on sample clustering, since it requires the calculation of distances between all the samples and all the cluster centroids. Let $N$ be the number of input samples, $D$ be the dimensionality of each sample and $K$ be the number of clusters that samples are partitioned into. If sample clustering is executed in a serial way, its time complexity will be O($N \cdot K \cdot D$). To reduce the time complexity, clustering for different samples can be executed in parallel, since the clustering process is independent for each sample. If the input sample set is divided into $M$ parts and the clustering tasks for the $M$ parts are executed concurrently, the time complexity for the sample clustering step can be reduced to O($N \cdot K \cdot D / M$). In addition, parallelism can also be exploited in the cluster centroids updating step, because samples in different parts can be accumulated concurrently. The intrinsic parallelism in the $k$-means algorithm discussed above make it suitable for the Map-Reduce programming model.

Map-Reduce is a powerful programming model for parallel computing. The model expresses the computation as Map and Reduce functions [13]. Instances of the Map function, called Mappers, operate individually and generate intermediate results, then the results are grouped and processed by instances of the Reduce function, called Reducers. The input task is segmented and executed as Map-Reduce jobs, which are allocated to the Mappers and Reducers. During the execution, enormous opportunities for parallelism are provided.

In this paper, a simplified Map-Reduce architecture is proposed to implement the $k$-means algorithm on an FPGA. The architecture consists of $M$ Mappers and one Reducer, as shown in the Figure 1. Each iteration in the $k$-means algorithm is executed as a Map-Reduce job. The performance of each Map-Reduce job is optimized by algorithmic segmentation, data path elaboration and automatic control. In addition, 32-bit single-precision float-point arithmetic is used to enhance the precision of calculation and the accuracy of clustering.

## 3.1 Algorithmic Segmentation

Each iteration in the $k$-means algorithm, a Map-Reduce job in our design, is segmented into the Map phase and the Reduce phase. The Map phase is responsible for sample clustering and accumulation, which are executed in the $M$ Mappers. The Reduce phase mainly takes charge of generating new cluster centroids by division and is executed in the single Reducer. Compared with the architecture proposed in [1], accumulations of samples in each cluster are offloaded to the Map phase rather than the Reduce phase. This helps reduce latency caused by data transfer, since samples are no longer required by the Reducer. Figure 1 shows the job distribution between the $M$ Mappers and the single Reducer. Each Map-Reduce job is executed in steps as follow:

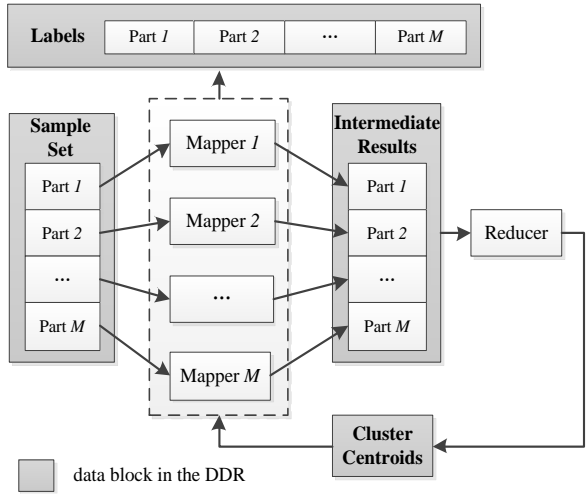

**Figure 1. Algorithmic Segmentation and Data Path**

(a) The cluster centroids are sent into each Mapper and the sample set is divided into $M$ parts. Then each part of the sample set is dispatched into a specific Mapper.
(b) Each Mapper clusters the input samples by distance calculation and comparison, then accumulates the samples in each cluster. Each Mapper generates two outputs: one consists of labels indicating the cluster that each sample is assigned to, while the other is comprised of intermediate results, including the number and the partial sums of samples in each cluster.
(c) The intermediate results from Mappers are grouped and then sent into the Reducer. The Reducer accumulates the numbers and the partial sums of samples for each cluster, then generates new cluster centroids by calculating the mean of samples in each cluster. The new cluster centroids will override the old ones and be used in the next iteration.

In this way, the intrinsic parallelism of the $k$-means algorithm is fully exploited and the runtime required for sample clustering and accumulation is reduced by a factor of $M$. The time complexity of the Map phase and the Reduce phase is $O(N \cdot K \cdot D / M)$ and $O(M \cdot K \cdot D)$ respectively. Due to the limitation of hardware resource, $M$ is much smaller than $N$, hence the Map phase is responsible for the majority of the total runtime when $N$ grows large. Additionally, $M$ can be configured to balance the system performance, hardware resource and memory bandwidth.

## 3.2 Data Path

In the proposed Map-Reduce architecture, data are transfered between the FPGA and the memory. Data copy in the memory, which is time-consuming, is avoided by elaborating the memory space allocation. In addition, the Xilinx Advanced eXtensible Interface (AXI) protocol [14] is adopted for high-performance data transfer.

Figure 1 shows the data interaction between the memory and the FPGA, as well as the memory space allocation. Samples in the sample set are stored continuously in the memory and equally divided into $M$ parts for $M$ Mappers. The intermediate results generated by Mappers are stored continuously in the memory, for the convenience of transferring through a Direct Memory Access (DMA) toward the Reducer later. The input cluster centroids for Mappers and the new cluster centroids generated by the Reducer share the same memory space. The new cluster centroids override the old one, and are used by Mappers in the next iteration. In this way, data copy in the memory is avoided, which largely reduces latency.

Two AXI4 interfaces, AXI4-Master and AXI4-Stream, are adopted for high-speed data transfer. The AXI4-Master interface is memory-mapped and allows high-performance burst transmission of up to 256 data transfer cycles with a single address phase. The AXI4-Stream interface is for high-speed streaming data transmission without address phases and allows unlimited data burst size [14]. When an AXI4-Stream interface is applied for data transfer between the memory and the FPGA, a DMA engine is required. Briefly, the AXI4-Master interface is commonly used for high-performance data transfer in relatively small amount, while the AXI4-Stream interface is suitable for high-speed data transmission in large scale.

In the proposed architecture, the space complexities of the sample input and the label output of each Mapper are $O(N \cdot D / M)$ and $O(N / M)$ respectively. These complexities become high for large sample set. In contrast, the space complexities of the cluster centroids input and the intermediate results output of each Mapper, along with the new cluster centroids output of the Reducer, are all $O(K \cdot D)$ and relatively low. Hence AXI4-Stream interfaces are applied to transfer samples and labels to ensure high performance, while AXI4-Master interfaces are used to transmit cluster centroids and intermediate results to reduce the consumption of DMAs. Besides, the space complexity of the input of the Reducer is $O(M \cdot K \cdot D)$, which means that the data volume is considerable when $M$ or $D$ grows large. Therefore, an AXI4-Stream interface is applied to avoid the bottleneck caused by data transmission.

## 3.3 Automatic Control

In traditional Map-Reduce frameworks, a master node is required to schedule the operation of Mappers and Reducers and to control the data transfer. Typically in FPGA-based implementations, the master node is a host CPU with programs for scheduling and control. The communication overhead between the host and the FPGA may reduce the performance of the entire system significantly. Therefore, the host programs are implemented by hardware circuits in the proposed Map-Reduce architecture. In addition, at the end of each iteration in the $k$-means algorithm, the system should adjudicate on whether the algorithm is completed according to the distortion error. This arbitration is also implemented on hardware. Details for automatic control are discussed in Sections 4.3 and 4.4.

## 4. K-MEANS IMPLEMENTATION

To implement the Map-Reduce architecture described in Section 3, four IP cores: *Kmeans_Map*, *Kmeans_Reduce*, *DmaScheduler* and *Iteration_Controller* are designed. The first two IP cores implement the function of the Mapper and the Reducer respectively. The *DmaScheduler* is designed to control the data transfer between the memory and the FPGA. The *Iteration_Controller* is used to control the execution of the $k$-means algorithm. The *Kmeans_Map* and the *Kmeans_Reduce* IP cores are described in C and synthesized by high level synthesis tool Vivado HLS provided by Xilinx [15], while the *DmaScehduler* and the *Iteration_Controller* are described with Hardware Description Languages (HDLs). Details for implementations of the four IP cores are discussed as follows.

```
1. function Kmeans_Map (Centroids_in, Sample_in,
       Mediate_out, Label_out)
2. #pragma HLS INTERFACE m_axi port=Centroids_in
3. #pragma HLS INTERFACE m_axi port=Mediate_out
4. #pragma HLS INTERFACE axis port=Sample_in
5. #pragma HLS INTERFACE axis port=Label_out
6. Initialization
7. Cache Centroids_in on FPGA as cluster centroids
8. for each sample from Sample_in
9. #pragma HLS PIPELINE
10.   Cache the sample on FPGA
11.   for each cluster centroid
12. #pragma HLS ARRAY_PARTITION
13.     Calculate distance between the sample and centroid
14.     Update the minimum distance and nearest cluster
15.   end for
16.   Assign the sample to the nearest cluster
17.   Update the counter in the correspondent cluster
18.   Update the accumulator in the correspondent cluster
19.   Update the distortion error by the minimum distance
20.   Output the label of the sample to DDR as Label_out
21. end for
22. Output intermediate results to DDR as Mediate_out
23. endfunciton
```

**Figure 2. Pseudo-code for *Kmeans_Map* IP Core**

## 4.1 Implementation of Mapper

Traditionally, IP cores are designed with HDLs such as VHDL or Verilog, which can be time-consuming tasks for complex algorithmic implementations. Applying HLS technique allows users to describe their designs in a higher abstraction level with advanced languages such as C/C++, thus largely reduces the complexity and cycles for development. The HLS tool can translate the descriptions from a high level to the register transfer level and automatically schedule the timing. In addition, pragmas are provided by the HLS tool for users to explore the parallelism in their implementations conveniently. The *Kmeans_Map* and *Kmeans_Reduce* IP cores are developed with HLS technique.

The *Kmeans_Map* IP core implements the functionality of the Mapper in proposed Map-Reduce architecture. Figure 2 shows the pseudo-code for describing the behavior of the *Kmeans_Map* IP core. Interfaces for inputs and outputs are specified in Lines 2-5, according to the consideration of data volume discussed in Section 3.2. In our implementation, samples are stored in the DDR rather than the on-chip memory on FPGA due to the relative small size of the on-chip memory. At the beginning of the execution, cluster centroids are first loaded from the DDR (Line 7). Then samples are streamed into the FPGA sequentially (Line 10). The cluster centroids and the sample in process are frequently accessed, hence they are cached in the local memory of the FPGA to reduce transmission latency. Subsequently, distances between the sample in process and all the cluster centroids are calculated according to the Euclidean metrics. Meanwhile, the present minimum distance and its corresponding ordinal of cluster are updated (Lines 11-15). Once the minimum distance is found, the sample is attached with a label indicating the cluster it belongs to, and the accumulator and counter in the corresponding cluster, along with the distortion error, are updated in the meantime (Lines 16-19). The label is then sent to the DDR (Line 20). When the processes of all the samples are completed, those intermediate results are sent to the DDR and then transferred into the *Kmeans_Reduce* IP core (Line 22).

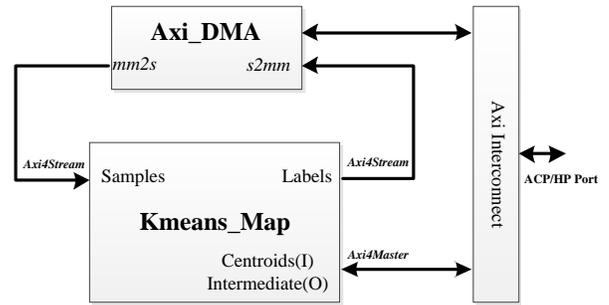

**Figure 3. Structure of Mapper Unit**

Two pragmas are applied in the *Kmeans_Map* IP core design for further optimization. In line 9, a *PIPELINE* pragma is applied to guide the HLS tool to construct pipelines automatically, which helps to reduce the initialization intervals between the processes of two consecutive samples by allowing concurrent executions [13]. In addition, the *ARRAY_PARTITION* pragma in line 12 is used to instruct the HLS tool to map the arrays in C code to registers rather than Block RAMs (BRAMs), which removes the potential bottlenecks caused by BRAM accessions. After high level synthesis, the initialization interval is equal to the dimensionality of each sample, since each dimension of a sample has to be accessed sequentially due to the feature of the AXI4-Stream interface. Besides, each *Kmeans_Map* IP core can achieve a throughput of 2.9 Gbps, according to the synthesis report provided by the HLS tool.

Each *Kmeans_Map* IP core, combined with an *Axi_DMA* IP core [16], constitutes a Mapper Unit, as shown in the Figure 3. The *Axi_DMA* IP core is a DMA engine encapsulated by AXI4 interfaces. As discussed in Section 3.2, the sample input and the label output are implemented as AXI4-Stream interfaces so that data transfer through a DMA engine. In contrast, the cluster centroids and the intermediate results share an AXI4-Master interface for transmission.

```
1. function Kmeans_Reduce (Mediate_in, Centroids_out)
2. #pragma HLS INTERFACE axis port=Mediate_in
3. #pragma HLS INTERFACE m_axi port=Centroids_out
4. Initialization
5. for intermediate results of each mapper from Mediate_in
6.   Update the counter in the correspondent cluster
7.   Update the accumulator in the correspondent cluster
8.   Accumulate the distortion error
9. end for
10. for each cluster
11. #pragma HLS PIPELINE
12.   Calculate the mean of samples in the cluster
13.   Move the cluster centroid to the mean point
14. end for
15. Output new centroids to DDR as Centroids_out
16. Compare distortion error with its value in last iteration
17. if change of distortion error is within a given threshold
18.   assert the iteration_done signal
19. endif
20. endfunction
```

**Figure 4. Pseudo-code for *Kmeans_Reduce* IP Core**

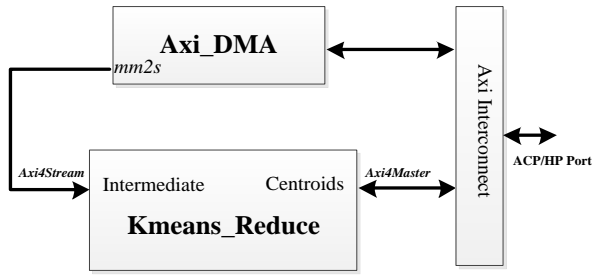

**Figure 5. Structure of Reducer Unit**

### 4.2 Implementation of Reducer

The *Kmeans_Reduce* IP core implements the functionality of the Reducer. The pseudo-code for describing the behavior of the *Kmeans_Reduce* IP core is shown in the Figure 4. Interfaces for the input and output are specified in Lines 2-3 according to the consideration of data volume discussed in Section 3.2. The execution proceeds in three steps. First, the number and the partial sums of samples in each cluster, along with the partial sum of the distortion error, are accumulated respectively, all of which are generated by Mappers as intermediate results (Lines 5-9). Then the mean of samples in each cluster is calculated by division to generate the new cluster centroids (Lines 10-14). A *PIPELINE* pragma is applied to allow the division to be executed in parallel to reduce latency (Line 11). Later, the new cluster centroids are sent to the DDR for use in the next iteration (Line 15) and the distortion error is compared to its value in the last iteration meanwhile (Line 16). If the change of the distortion error is within a user-defined error threshold, a one-bit signal *iteration_done* is asserted, which indicates that the *k*-means algorithm is done and the final clustering results are ready (Lines 17-19).

The *Kmeans_Reduce* IP core, combined with an *Axi_DMA* IP core, constitutes the Reducer Unit, as shown in the Figure 5. According to the discussion in Section 3.2, an AXI4-Stream interface is applied to the intermediate results input so that data transfer through a DMA engine. In contrast, an AXI4-Master interface is applied to the new cluster centroids output.

Both of the *Kmeans_Map* and *Kmeans_Reduce* IP cores are flexible and scalable. Parameters such as the number of samples that one IP core can process each time, the dimensionality of each sample and the number of cluster that the sample set is partitioned into can be configured in a simple way, allowing users to customize their designs for particular applications conveniently.

### 4.3 Hardware Integration

The proposed Map-Reduce architecture for the *k*-means algorithm is implemented on a Zynq device provided by Xilinx [16], while it can be easily migrated to other hardware platform. Figure 6 shows the hardware architecture on the FPGA. The system is divided into three hierarchies. First, as mentioned in Sections 4.1 and 4.2, each *Kmeans_Map* IP core and a DMA engine form a Mapper Unit, while the single *Kmeans_Reduce* IP core and a DMA engine form the Reducer Unit. Then the *M* Mapper Units constitutes the Mapper Block with a *DmaScheduler*, while the Reducer Unit and a *DmaScheduler* constitute the Reducer Block. On the top hierarchy, the Mapper Block and the Reducer Block, along with the *Iteration_Controller*, make up the entire system.

Implementations of the *Kmeans_Map* and the *Kmeans_Reduce* IP core have been explained in Sections 4.1 and 4.2. The other two IP cores *Iteration_Controller* and *DmaScheduler* are designed for automatic control. As mentioned in Section 3.3, a master node, typically a host with programs for scheduling and control, is required in traditional Map-Reduce frameworks. In contrast, the host programs are implemented by hardware circuits in the proposed architecture, which helps reduce the communication overhead between the host and the FPGA. The principle of the *Interation_Controller* is explained as follows, while the mechanism of the *DmaScheduler* is discussed in Section 4.4.

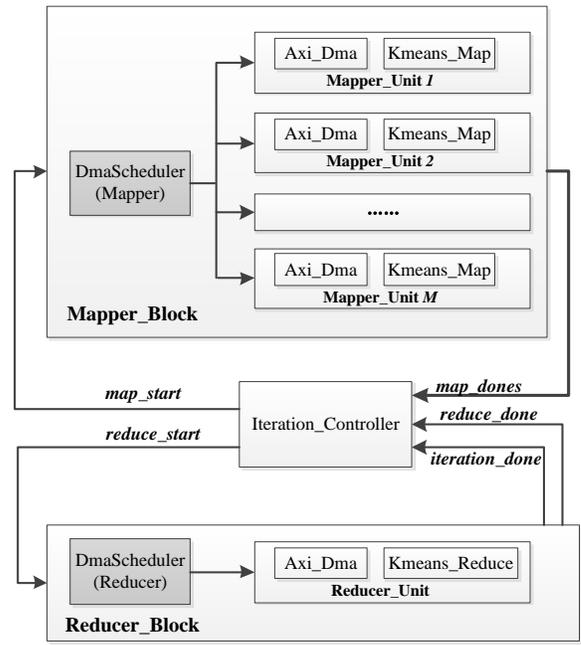

**Figure 6. Hardware Architecture on FPGA**

In our system, to maintain the synchronization of data and avoid malfunction, only when all the Mapper Units complete their clustering tasks should the Reducer Unit starts to work, and only when the Reducer Unit completes the update of cluster centroids should a new iteration begins. Therefore, a controller is designed to schedule the operations of the Mapper Units and the Reducer Unit. Under the control of the *Iteration_Controller*, the hardware system operates in steps as follow:

(a) The *Iteration_Controller* generates a *map_start* signal and the *DmaScheduler* in the Mapper Block begins to work.
(b) The *DmaScheduler* in the Mapper Block starts the DMA engines in all the Mapper Units and the Mapper Units begin to work.
(c) When a Mapper Units complete its task, a *map_done* signal is sent to the *Iteration_Controller*. Once the *Iteration_Controller* receives *M map_done* signals, it generates a *reduce_start* signal, notifying the *DmaScheduler* in the Reducer Block to work.
(d) The *DmaScheduler* in the Reducer Block starts the DMA engines in the Reducer Unit and the Reducer Unit begins to work.
(e) When the Reducer Unit completes its task, a *reduce_done* signal is sent to the *Iteration_Controller*, which means an iteration of the *k*-means algorithm is done.
(f) If the *Iteration_Controller* receives an *iteration_done* signal generated by the Reducer Unit, the iteration stops because the final clustering results are ready. Otherwise, steps (a)-(e) are repeated.

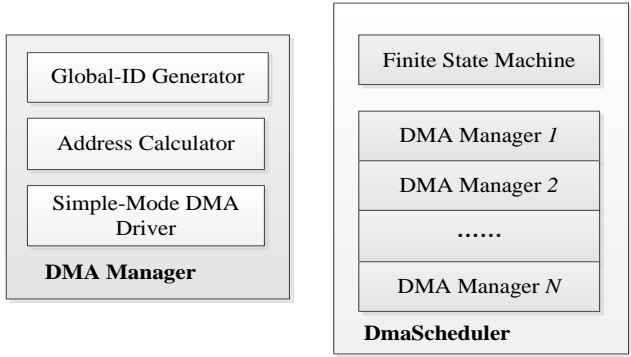

**Figure 7. Structure of *DmaScheduler***

## 4.4 DmaScheduler

The *DmaScheduler* is designed to automatically control the data transmission by scheduling the operation of the DMA engines in the system. As shown in the Figure 6, two *DmaSchedulers* are instantiated, one in the Mapper Block while the other in the Reducer Block. The structure of the *DmaScheduler* is shown in the Figure 7. Each *DmaScheduler* consists of several DMA managers and a finite state machine.

As discussed in Sections 4.1 and 4.2, an *Axi_DMA* IP core is instantiated for data transfer through the AXI4-Stream interface between the FPGA and the DDR in each Mapper Unit and the Reducer Unit. The operation of these DMA engines needs to be scheduled systematically. In each *DmaScheduler*, the DMA Managers are designed to control the behaviors of the DMA engines. Each DMA Manager is responsible for one DMA engine and consists of a Global-ID Generator, an Address Calculator and a Simple-Mode DMA Driver. Before the data transfer, the DMA engine needs to acquire the start address of the data block to be transferred. The start address is generated by the Address Calculator. As stated in Section 3.2, data in data blocks including the sample set, intermediate results and labels are stored continuously. Hence the start addresses of these data blocks can be calculated easily according to their specific IDs, which are assigned to each data block by the Global-ID Generator. The start address for each data block is formulated as a sum:

$$\text{Address} = \text{BaseAddr} + \text{ID} \times \text{Length} \quad (1)$$

Besides, the Simple-Mode DMA Driver in each DMA Manager is used to manage the behaviors of the DMA engine automatically. In the implementation, each DMA engine operates in Simple Mode [16]. The Simple Mode allows a DMA engine to be configured and used easily, but only supports up to 8MB of data volume for each data transmission. For volume larger than 8MB, data can be partitioned and transferred successively.

In addition, the finite state machine in the *DmaScheduler* is designed to control the operation of the IP core. Taking the *DmaScheduler* in the Mapper Block as an example, the operation is divided by states as follows:

(a) When a *map_start* signal is received from the *Iteration_Controller*, all the DMA Managers begin to generate IDs for data blocks, calculate their start addresses and start the DMA engines in the Mapper Units.
(b) Samples are fetched by DMA engines from the DDR and streamed into the *Kmeans_Map* IP cores.
(c) When the *Kmeans_Map* IP cores complete the clustering tasks, labels are streamed out to the DDR through the DMA engines.
(d) The *DmaScheduler* stops its operation when all DMA Managers inside complete their jobs, then it waits for the next *map_start* signal.

## 5. EXPERIMENTAL RESULTS

The proposed design is evaluated on an Xilinx ZC706 FPGA board with an xc7z045ffg900-2 FPGA. In evaluation, the input sample set is fetched from the host through the PCIe ×4 gen2 interface and the transmission bandwidth can achieve 7.6 Gbps in our implementation. Since the transfer of samples from host to FPGA happens only once for each *k*-means execution and the overhead is small compared to the total runtime, this overhead is neglected in the following evaluation. Besides, to avoid the bottleneck caused by the memory bandwidth, two on-board DDRs are utilized: the DDR in the Processing System and the DDR in Programmable Logic [4]. The system runs at 100MHz for evaluation, while a higher frequency can be achieved.

The evaluation is divided into three parts. In section 5.1, performance of our implementation is evaluated and compared with other FPGA-based implementations. Then effects caused by variations of the parameters (*M* and *N*) are explored in Section 5.2. Finally, the resource utilization is discussed in Section 5.3.

## 5.1 Performance Evaluation and Comparison

The performance of our implementation is measured by runtime and throughput. For better comparison, the same sample set as in [1] is used for evaluation. The sample set is a power consumption data set from the UCI Machine Learning Repository [18]. It records the electric power consumption in one household with a one-minute sampling rate over 4 years and contains 2075259 samples with 9 attributes each. Two input sample sets are created by extracting attributes from the original data set. One consists of 2-D samples, whose data are extracted from attributes *global_active_power* and *global_reactive_power*. The other makes up of 4-D samples, whose data are extracted from attributes *global_active_power*, *sub_metering_1*, *sub_metering_2* and *sub_metering_3*. Besides, the initial cluster centroids are generated randomly in software, and 12 Mappers are used due to the limitation of hardware resource.

The evaluation results are shown in the Table 1. It shows that the 2-D and the 4-D sample sets converge after 34 and 11 iterations on average respectively when partitioned into 4 clusters. The runtime required for one iteration is similar for the two sample sets, which means that the throughput for each iteration of the 4-D sample set is twice the throughput of the 2-D sample set. Additionally, a throughput of 28.74 Gbps for one iteration is achieved for the 4-D sample set.

The performance is compared with that of four representative FPGA-based implementations for *k*-means acceleration. The first one is a conventional implementation for clustering microarrays [9]. The second one is an optimized implementation based on the kd-tree data structures [11]. The third one is an implementation that follows the Map-Reduce model and targets for big data applications [1]. And the last one is an OpenCL-based implementation. For comparison, the throughput of a single iteration in the $k$-means algorithm is used as the indicator and can be calculated according to the equation below. $W$ is the bitwidth of each dimension of a sample:

$$\text{Throughput} = \frac{N \times D \times W}{Rumtime} \; bit/s \quad (2)$$

**Table 1. Performance Evaluation**

| D | Average number of iteration | Single iteration time | Single iteration Throughput |
|---|---|---|---|
| 2 | 34 | 8.48 ms | 14.40 Gbps |
| 4 | 11 | 8.50 ms | 28.74 Gbps |

Table 2. Performance Comparisons with Existing Works

| Implementation | Throughput (Gbps) | Performance Speedup |
|---|---|---|
| Hussain's [9] (Xilinx 4vfx12) | 1.11 | 25.89x |
| Winterstein's [11] (Xilinx 7vx485t-1) | 7.32 | 3.93x |
| Choi's [1] (Xilinx 7k325t-2) | 0.55 | 52.25x |
| Ramanathan's [12] (Altera Stratix V D5) | 5.27 | 5.45x |
| This paper (Xilinx 7z045-2) | 28.74 | 1x |

Comparison results are shown in the Table 2. It can be seen that the proposed implementation provides at least 3.93x speedup over the other four implementations. Moreover, the first two implementations use fixed-point arithmetic, while 32-bit single-precision float-point arithmetic is applied in our implementation. Hence our implementation achieves high performance when maintains high precision in the meantime.

Additionally, the benefit brought by applying the automatic control on hardware is evaluated. The runtime for one iteration is reduced by 1.2% after integrating the *Iteration_Controller* and the *DmaScheduler* into the Map-Reduce architecture. Since the runtime can grow large for large-scale inputs, the 1.2% performance enhancement for each iteration is not negligible. Furthermore, the workload of the host CPU is reduced and the control procedures are simplified by applying automatic control.

## 5.2 Effects of Parameters

As mentioned in Section 3.1, the number of Mappers ($M$) can be configured to balance the system performance, hardware resource and memory bandwidth. Adding Mappers can enhance performance by increasing parallelism, but consumes more resources of an FPGA in the meantime. Thus in this section, the effect on performance caused by adding Mappers is explored.

The sample set used for evaluation is generated manually on Matlab, which consists of 512000 4-D samples. Figure 8 displays the throughput of one iteration as $M$ increases. It can be seen that the throughput increases in an approximately linear behavior with the increase of $M$, which means that the performance of the system benefits from the extra parallelism brought by adding Mappers. Besides, the growth of throughput slows down slightly when $M$ grows larger than 8. This is caused by the sharing of a single High Performance (HP) transmission port [17] by two or more Mappers.

Additionally, the performance is evaluated as input samples increase, especially under large-scale inputs. A series of 4-D

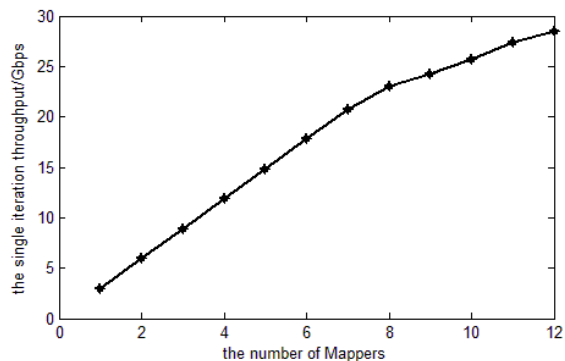

Figure 8. Throughput as *M* Varies

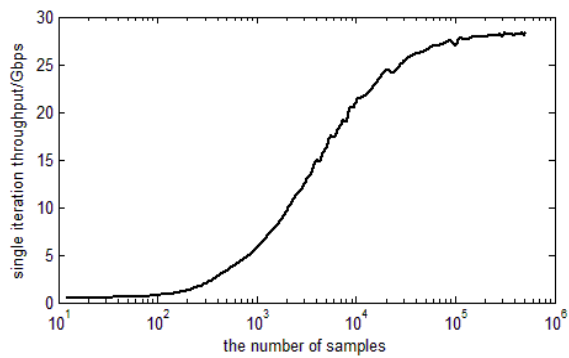

Figure 9. Throughput as *N* Varies

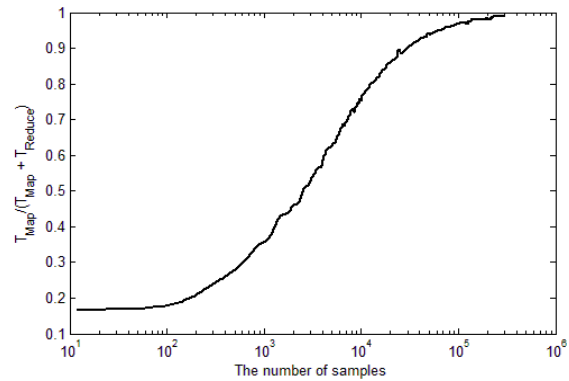

Figure 10. Ratio of Runtime

sample sets with different data volumes are generated for evaluation. 12 Mappers are used in this evaluation and results are shown in the Figure 9. It can be seen that the throughput of one iteration increases rapidly under small inputs ($N < 10^4$), and remains high and stable under large-scale inputs ($N > 10^5$). This means that the proposed architecture is applicable for big data applications.

Moreover, the ratio between the runtime of the Map phase and the total runtime is measured as $N$ varies. As shown in the Figure 10, the ratio increases rapidly when the input is small, while the ratio tends to remain high and constant under large inputs. When $N$ grows larger than 10000, the Map phase is responsible for more than 85% of the total runtime. This meets the theoretical prediction in Section 3.1: The Map phase occupies most of the total runtime when $N$ grows large, since the runtime for the Map phase is approximately proportional to $N$, whereas the variation of $N$ has no effects on the Reduce phase.

**Table 3. Resource Utilization (M = 12)**

| FPGA Resource | Amount | Utilization Ratio |
|---|---|---|
| Slice FF | 208152 | 47.61% |
| Slice LUTs | 178185 | 81.51% |
| Block RAM Tile | 159.5 | 29.27% |
| DSP | 412 | 45.78% |

## 5.3 Resource Utilization

Resource utilization of the FPGA for implementing the proposed Map-Reduce architecture is also evaluated. As shown in the Table 3, the implementation consumes more than 80% of LUTs and nearly half of FFs and DSPs on the xc7z045ffg900-2 FPGA when 12 Mappers are used. The utilization of hardware resource on the FPGA is maximized to achieve the highest performance.

## 6. CONCLUSION

In this paper, a simplified Map-Reduce architecture is proposed to implement the *k*-means algorithm on an FPGA. Algorithmic segmentation, data path elaboration and automatic control are applied to optimize the performance of each iteration in the *k*-means algorithm. In addition, HLS technique is utilized to reduce the development complexity and cycles. And float-point arithmetic is used to increase the accuracy of clustering. During evaluation, the implementation shows a throughput of 28.74 Gbps for one iteration in the *k*-means algorithm and a speedup of at least 3.93x over four representative existing implementations. The performance can satisfy the high computational requirement in big data applications. Future work includes extending the proposed architecture to multi-FPGA implementations and other data mining algorithms.